\documentclass{elsarticle}
\biboptions{sort&compress}
\usepackage[utf8]{inputenc}
\usepackage[T1]{fontenc}

\usepackage[colorlinks]{hyperref}
\usepackage{lineno}

\usepackage{IEEEtrantools}
\usepackage{mathtools}
\usepackage{ae,aecompl}
\usepackage{graphicx}
\usepackage{floatrow}
\usepackage{amsmath}
\usepackage{amssymb}
\usepackage{color}
\usepackage{dcolumn,booktabs}
\usepackage{siunitx}
\usepackage{verbatim}
\usepackage{rotating}
\DeclareUnicodeCharacter{00A0}{~}
\graphicspath{{pics/}}

\newcommand{\expect}[1]{\ensuremath{\left\langle \, #1 \, \right\rangle}}

\journal{International Journal of ...}

\begin{document}

\begin{frontmatter}

\title{Flow tracing as a tool set for the analysis of networked large-scale renewable electricity systems}


\author[fias]{Jonas~H\"{o}rsch\corref{corrauthor}}
\cortext[mycorrespondingauthor]{Corresponding author}
\ead{hoersch@fias.uni-frankfurt.de}

\author[aarhus,fias]{Mirko~Sch\"{a}fer}
\author[kasseluni,kasseliwes]{Sarah~Becker}
\author[fias]{Stefan~Schramm}
\author[aarhus]{Martin~Greiner}

\address[fias]{Frankfurt Institute for Advanced Studies, 60438~Frankfurt~am~Main, Germany}
\address[aarhus]{Department of Engineering, Aarhus University, 8000~Aarhus~C, Denmark}
\address[kasseluni]{Department of Electrical Engineering and Computer Science, Kassel University, 34125~Kassel}
\address[kasseliwes]{Fraunhofer IWES, 34119~Kassel, Germany}

\begin{abstract}
  The method of flow tracing follows the power flow from net-generating sources
  through the network to the net-consuming sinks, which allows to assign the
  usage of the underlying transmission infrastructure to the system
  participants. This article presents a reformulation that is applicable to
  arbitrary compositions of inflow appearing naturally in models of large-scale
  electricity systems with a high share of renewable power generation. We
  propose an application which allows to associate power flows on the grid to
  specific regions or generation technologies, and emphasizes the capability of
  this technique to disentangle the spatio-temporal patterns of physical imports
  and exports occurring in such systems. The analytical potential of this method
  is showcased for a scenario based on the IEEE 118 bus network.
\end{abstract}


\end{frontmatter}

\section*{Nomenclature}
\addcontentsline{toc}{section}{Nomenclature}
\subsection*{Indices and Labels}
\begin{IEEEdescription}[\IEEEusemathlabelsep\IEEEsetlabelwidth{$\alpha, \beta, \tau$}]
\item[$n, m, k$] Index of buses.
\item[$l, l'$] Index of lines.
\item[$\alpha, \beta, \tau$] Labels of regions and technologies for
  grouping the power injection and flows.
\end{IEEEdescription}

\subsection*{Constants, Variables and Functions}
\begin{IEEEdescription}[\IEEEusemathlabelsep\IEEEsetlabelwidth{$p_l(q_{l,\alpha}|F_l)$}]
\item[$P_n(t)$] Net power injection at bus $n$ (MW).
\item[$G^{\tau}_n(t)$] Power generation by technology $\tau$ at bus $n$ (MW).
\item[$L_n(t)$] Load at bus $n$ (MW).
\item[$F^{out}_{n\to m}(t)$] Power outflow from bus $n$ in direction of bus $m$ (MW).
\item[$F^{in}_{n\to m}(t)$] Power inflow to bus $m$ from bus $n$ (MW).
\item[$F_l(t)$] Absolute value of the power flow on line $l$.
\item[$\chi_{n\to m}(t)$] Loss in the transmission line between bus $n$ and $m$ (MW).
\item[$q^{in}_{n,\alpha}(t)$] In-partition, the share of the injected power at bus
  $n$ attributed to owner $\alpha$.
\item[$q^{out}_{n,\alpha}(t)$] Out-partition, the share of the consumed power at
  bus $n$ attributed to owner $\alpha$.
\item[$q_{l,\alpha}(t)$] Line-flow partition, the share of the power flow through
  line $l$ attributed to owner $\alpha$.
\item[$p_l(F_l)$] Probability for a flow $F_l$ on line $l$.
\item[$p_l(q_{l,\alpha}|F_l)$] Conditional probability for a share $q_{l,\alpha}$
  of owner $\alpha$ in case of a flow $F_l$.
\item[$h_{l,\alpha}(F_l)$] Average share of owner $\alpha$ on the link $l$ for a
  flow $F_l$.
\item[$w_{l,\alpha}(\mathcal{K})$] Weight for the usage of the capacity increment
  between $\mathcal{K}$ and $d\mathcal{K}$ attributed to owner $\alpha$ on the
  link $l$.
\item[$\mathcal{K}^T$] Transmission capacity of the network (MW).
\item[$\mathcal{K}^T_l$] Transmission capacity of line $l$ (MW).
\item[$\bar{\mathcal{K}}^T$] Transmission capacity of the network including length (MW km).
\item[$\bar{L_l}$] Length of transmission line $l$ (km).
\item[$\bar{D}_n$] Average graph distance of bus $n$ (km).
\item[$\mathcal{M}^{(1 \dots 4)}_{\alpha,\tau}$] Transmission network usage measures (MW km).
\end{IEEEdescription}
\section{Introduction}
The electricity system is built up of a complex interwoven network of
technologies, which provides the backbone for our modern society. In the past,
this network was characterized by power flows from large central power plants
downstream through the grid to the consumers, with only very limited
interactions between different geographical regions. Today, the rising share of
decentralized, fluctuating renewable generation and the increasing
inter-dependence of international electricity markets has led to a more
dynamical system: the power grid has become the underlying infrastructure for a
complex pattern of long-range power flows between a heterogeneous distribution
of power generation to consumers, integrating not only dispatchable conventional
generation, but also electricity from offshore wind farms, wind and solar parks
and roof-top solar panels. In this context, a deeper understanding of the
emerging power flow patterns is of paramount importance on different levels: For
instance, internationally integrated electricity markets need to incorporate
possible network congestion into their market design~\cite{aguado2012}, whereas
network expansion plans attempt to minimize this congestion in the long
run~\cite{tyndp,hagspiel2015}. Also the delevopment of fair and transparent grid
usage fee systems, or public discussions concerning the benefit of new
infrastructure projects rely strongly on insights concerning the composition and
dynamics of the flow pattern in the network~\cite{consentec06, Neukirch2016}.
In this article we present a reformulation of a well-known method of flow
allocation, denoted as average participation or flow tracing, that is well
adapted to the challenges of the system analysis of complex modern electricity
systems. Different approaches to the problem of flow allocation in power grids
are often derived from circuit theory \cite{conejo2007,chen2016} or are based on
approximations of the complex power flow equations for AC electrical networks
\cite{rudnick1995,brown2015}. For the application of such methods to the problem
of flow allocation in large-scale models of electricity systems, one has to
factor in the potentially coarse-grained nature of such models. Both the network
buses and transmission lines might be aggregated representations of lower level
infrastructures, which cannot be included in detail in the model due to
computational limitations or lack of data
\cite{jebaraj2006,connolly2010,hoersch2016}. The method of flow tracing can be
applied directly to the overall power flow pattern in the system, and thus does
not explicitly have to take into account the underlying modeling details. By
tracing what we term in-partitions, we show how the known composition of
network-injected power generation can be followed through the grid and thus be
transferred to the power flows and composition of net consumption at the sink
nodes. In this way the location of generation of power flow can be connected to
its location of consumption, thus disentangling the complex spatio-temporal
patterns of imports and exports inherent to interconnected electricity systems
with a high share of renewable generation. We showcase the potential of this
methodological tool set by application to the Scenario 2023B of the IEEE 118-bus
model adapted by Barrios et al. at RWTH Aachen with renewable generation
capacities and hourly availability for a model year as a benchmark for
transmission expansion algorithms~\cite{barrios2015}.

After a short review of flow tracing, Sec.~\ref{sec:methodology} introduces the
reformulated flow tracing technique and a measure of network usage. The
subsequent Sec.~\ref{sec:application} showcases two exemplary applications:
Firstly the tracing of power flow of different generation types between several
regions across a network model based on the IEEE 118 bus case, and secondly a
comparison of a statistical transmission capacity usage measure with several
alternative allocation mechanisms. Section~\ref{sec:concl} concludes the paper.


\section{Methodology}
\label{sec:methodology}

Flow tracing was introduced as a loss-allocation scheme by Bialek et
al. based on solving linear equations~\cite{bialek1996} and in parallel
by Kirschen et al. as an analytical tool using a graph-based, iterative
approach~\cite{kirschen1997}.

It was soon after proposed as a transmission-usage allocation
scheme~\cite{bialek1997,bialek1998,strbac1998,kirschen1999}. Subsequently,
the method was discussed to cover concrete supplementary charge schemes
for cross-border trades~\cite{bialek2003,bialek2004-2}, in view of the
discussion about the mechanism of inter-transmission system operator
compensation in Europe~\cite{comillas02, consentec06, camacho2007}.

Of the other network-cost allocation methods -- reviewed in \cite{lima2009} or
\cite{abou2008}, for instance -- we only want to highlight marginal
participation~\cite{perezarriaga2000}, which attributes transmission capacity
according to linear sensitivities of network flows to differential bus
injections as captured by the power transfer distribution factors
(PTDF)~\cite{grainger1994-book}. Due to its influence on the PTDF, for this
method the choice of the slack bus has to be taken into account
explicitly~\cite{vazquez2002}, whereas for the flow tracing technique this
choice only affects the total power flow but not the allocation mechanism.


\subsection{Power flow}
\label{sec:power_flow}

The active power flow in an electricity system satisfies Kirchhoff's
current law. If the net power injection at bus \(n\) from
generators and loads is given by \(P_n\), and \(F^{in/out}_{n\to m}\) are the power in- and outflows from bus \(n\)
to \(m\), then the power flow through node \(n\) is conserved as
\begin{linenomath}
\begin{equation}
  P^{in}_n + \sum_m F^{in}_{m\to n} = P^{out}_n + \sum_m F^{out}_{n\to m}~.
  \label{eq:power-conservation}
\end{equation}
\end{linenomath}
Here we use the positive and negative injections $P^{in}_n$ and $P^{out}_n$
at node \(n\) and invoke the convention that all \(F^{out}_{m\to n}\)
and \(F^{in}_{m\to n}\) are positive or zero.

Table~\ref{tab:eupower} introduces a particular snapshot in a simple network
with four buses with generation $G_n$, load $L_n$ and im-/exports $I_n$/$X_n$
with other buses not represented explicitly. In this example, we take the
positive injection as the net surplus between generation $G_n$ and demand $L_n$
plus the imports $I_n$, while the negative injection follows from the deficit
and exports $X_n$, as
\begin{linenomath}
\begin{IEEEeqnarray}{rCl"rCl}
  P^{in}_n &=& \max \{(G_n - L_n), 0\} + I_n~, &   P^{out}_n &=& \max \{-(G_n - L_n), 0\} + X_n~.
\end{IEEEeqnarray}
\end{linenomath}
The flows and line-losses are illustrated in Fig.\ref{fig:4-node-num}. The
convention means that the line from bus~1 to bus~3 is described by
\(F^{out}_{\mathtt{1}\to \mathtt{3}} = \SI{2.2}{GW}\), \(F^{in}_{\mathtt{1}\to
  \mathtt{3}} = \SI{1.8}{GW}\), \(F^{in}_{\mathtt{3}\to \mathtt{1}} = 0\) and
\(F^{out}_{\mathtt{3}\to \mathtt{1}} = 0\).

Here and in general the outflow from bus~\(n\) to~\(m\), \(F^{out}_{n\to m}\), is
larger than the inflow to~\(m\), \(F^{in}_{n \to m}\) due to losses in the
transmission line \(n \to m\). We denote them by \(\chi_{n \to m} = F^{out}_{n \to m}
- F^{in}_{n \to m}\).

\begin{table}
  \caption{Power generation and consumption of a simple four bus network with
    im-/exports with external buses in GW.}
  \label{tab:eupower}
  \centering
\begin{tabular}{llrrrrrrrr}
\toprule
$n$ & $G_n$ & \multicolumn{3}{l}{of which in \%} &  $L_n$ &  $G_n - L_n$ & $I_n$ & $X_n$ \\
    &       &  wind &  solar & other  &       &       &      &      \\
\midrule
1   &  76.0 &    16 &     19 &     65 &  65.5 &  10.5 &  0.9 &  5.6 \\
2   &  20.5 &     8 &      0 &     92 &  21.1 &  -0.6 &  0.9 &  0.6 \\
3   &   8.5 &     2 &     13 &     85 &   8.0 &   0.5 &  0.0 &  1.8 \\
4   &   7.3 &    12 &      6 &     82 &   7.5 &  -0.3 &  0.0 &  2.5 \\
\bottomrule
\end{tabular}
\end{table}

\subsection{Flow tracing}
\label{sec:flow_tracing}

\begin{figure}
  \floatbox[{\capbeside\thisfloatsetup{capbesideposition={left,bottom},capbesidewidth=0.4\linewidth}}]{figure}[\FBwidth]
  {\caption{Power flows and injections in the simple four bus network
      introduced in Table~\ref{tab:eupower} in units of
      GW. Injections are composed of generation, consumption, im-
      and exports as described in the text}\label{fig:4-node-num}}
  {\includegraphics[width=\linewidth,trim=1.5em 2em 2em 2em]{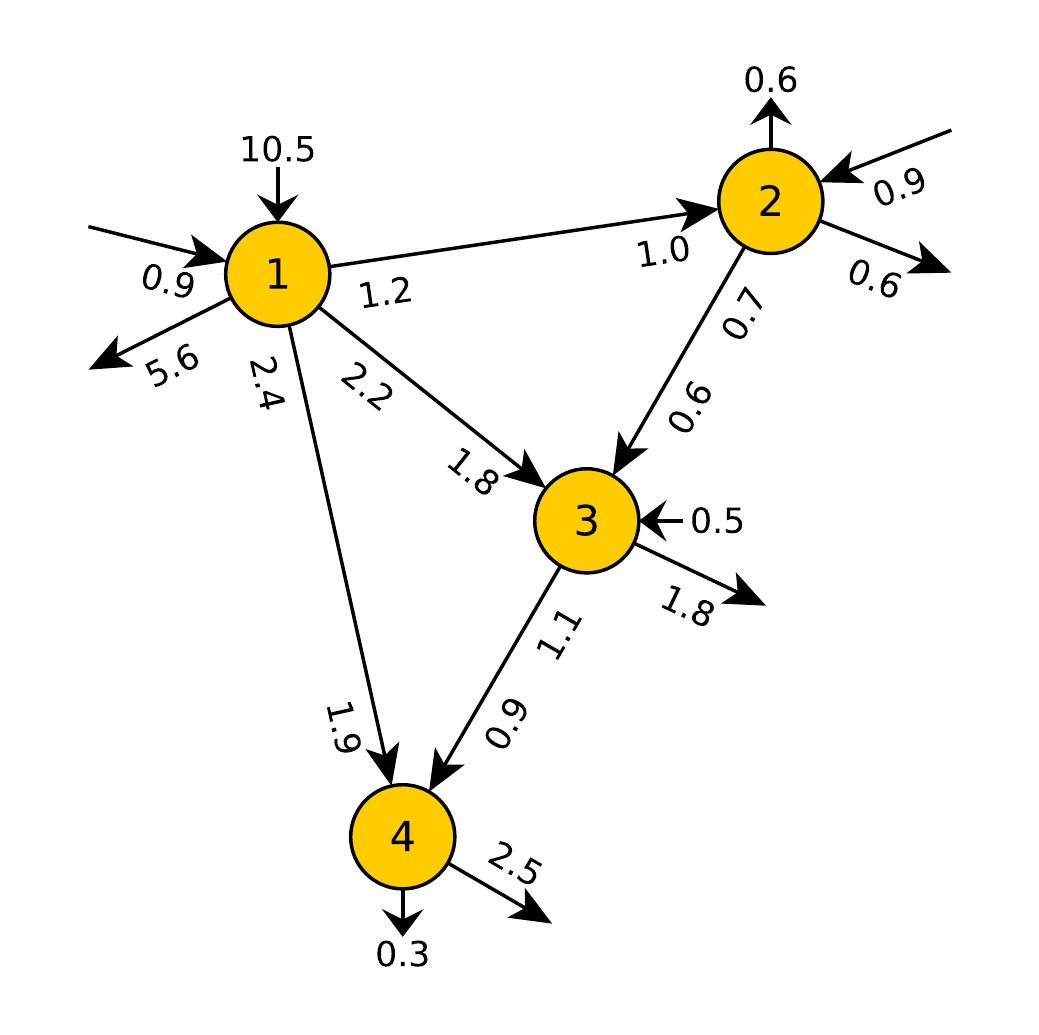}}
\end{figure}

The flow tracing method by Bialek and Kirschen~\cite{bialek1996,kirschen1997}
follows the power flow from individual buses through the network and decomposes
the flow on the power lines into contributions associated to each bus. Since for
large-scale electricity systems, the injection $P^{in}_{n}$, in general, already
contains several constituents, we introduce an in-partition $q^{in}_{n,\alpha}$
associating the power injection at each bus $n$ to a set of components
${\alpha}$. For the power flows of the four bus example, we will use the
components $\{1, 2, 3, 4, I\}$ with the in-partition
\begin{linenomath}
\begin{equation}
  q^{in}_{n,\alpha} = \left\{\begin{aligned}
    &\frac{\max\{G_n-L_n, 0\}}{P^{in}_{n}} & &\text{for } \alpha = n \wedge P^{in}_n > 0 ~, \\
    &\frac{I_n}{P^{in}_{n}}             & &\text{for } \alpha = I \wedge P^{in}_n > 0~, \\
    &0                             & &\text{else.}
  \end{aligned} \right\} =
  \begin{pmatrix}
    \frac{10.5}{11.4} & 0 & 0 & 0 & \frac{0.9}{11.4} \\
    0 & 0 & 0 & 0 &     1      \\
    0 & 0 & 1 & 0 &     0      \\
    0 & 0 & 0 & 0 &     0
  \end{pmatrix}
  \label{eq:qin-eu-num}
\end{equation}
\end{linenomath}
to differentiate the imports $I_n$ entering at each bus from the power
generated there. Note that the component $I$ is associated with injected power
throughout the network. Similarly, another in-partition for components
\{wind, solar, other, imports\} is able to
encode the relative shares of wind, solar and other energy generation sources
from Table~\ref{tab:eupower}.

Flow tracing follows the diffusion of the different components $\alpha$ by
assuming conservation of the partial power flows at bus $n$ in analogy
to~(\ref{eq:power-conservation})
\begin{linenomath}
\begin{equation}
  \label{eq:alpha_flow_conservation_separate}
  q_{n,\alpha}^{in} P_{n}^{in} + \sum_{m} q^{(m\to n)}_{m,\alpha} F^{in}_{m \to n} = q^{(out)}_{n,\alpha}P_{n}^{out} + \sum_{m} q^{(n\to m)}_{n,\alpha} F^{out}_{n\to m}~.
\end{equation}
\end{linenomath}
In general there is a degree of freedom in relating $q^{(out)}_{n,\alpha}$ and $q^{(n\to m)}_{n,\alpha}$ under the boundary condition of assuring conversation of partial flows. It
is nevertheless intuitive to assume that the power contributions mix perfectly
in each bus and the partitions of the flows leaving a bus are all identical
$q_{n,\alpha} = q^{(out)}_{n,\alpha} = q^{(n\to m)}_{n,\alpha}$. This assumption
equally underlies the tracing by Bialek and Kirschen and is known as
proportional sharing. Bialek~et.~al. were able to show the proportional sharing
principle to coincide with the Shapley value of a stylized game of loss
attribution played by two generators feeding into the same line~\cite{bialek2004},
but it can rightfully be contested for practical purposes. For the purpose of
flow allocation in large-scale electricity models, we suggest this realization
in particular due to its intuitiveness and lack of additional parameters. With
the proportional sharing assumption
Eq.~(\ref{eq:alpha_flow_conservation_separate}) reduces to a system of $N\times
A$ equations for $N\times A$ unknowns $q_{n,\alpha}$ 
\begin{linenomath}
\begin{equation}
  \label{eq:alpha_flow_conservation}
  q_{n,\alpha}^{in} P_{n}^{in} + \sum_{m} q_{m,\alpha} F^{in}_{m \to n} = q_{n,\alpha} \left( P_{n}^{out} + \sum_{m} F^{out}_{n\to m}\right)~,
\end{equation}
\end{linenomath}
with $A$ denoting the number of components $\alpha$. If we eliminate inert buses without any flows from the network
(without any loss of generality) and abbreviate the power leaving a
bus as nodal flow $F_n := P_{n}^{out} + \sum_{k}F^{out}_{n\to k}$, we
can rearrange Eq.~(\ref{eq:alpha_flow_conservation}) to
\begin{linenomath}
\begin{equation}
  \label{eq:flow_tracing_matrix}
q_{n,\alpha}^{in} P_{n}^{in} = F_{n} \sum_{m} \left[\delta_{n,m} - \frac{F^{in}_{m\to n}}{F_n}\right] q_{m,\alpha}
\end{equation}
\end{linenomath}
with the Kronecker delta \(\delta_{n,m}\). Finally, with the
definition of the matrix
\begin{linenomath}
\begin{equation}
  D_{n,m} = \frac{F^{in}_{m\to n}}{F_n}
\end{equation}
\end{linenomath}
capturing the share that the power from bus $m$ contributes to the
nodal flow through bus $n$, Eq.~(\ref{eq:flow_tracing_matrix}) can be
rendered in matrix notation as
\begin{linenomath}
\begin{equation}
  \label{eq:flow_tracing_matrix2}
  \mathrm{diag}(P^{in}) q^{in} = \mathrm{diag}(F) (1 - D) q~.
\end{equation}
\end{linenomath}
For a lossless power flow, $F^{out}_{n\to m} = F^{in}_{n\to m}$, $1-D$ is the transpose of
the downstream distribution matrix $A_d$ in Bialek's formulation and together
with $q^{in} = q = 1$ the proposed method reduces to Bialek's flow tracing. The
steps from Eq.~(\ref{eq:alpha_flow_conservation}) to
Eq.~(\ref{eq:flow_tracing_matrix2}) illustrate the equivalence of the two
formulations of flow tracing discussed in the literature as a linear algebra
problem \cite{bialek1996} and a graph-based algorithm~\cite{kirschen1997}.

Eq.~(\ref{eq:flow_tracing_matrix2}) is solved formally as
\begin{linenomath}
\begin{equation}
  \label{eq:flow-tracing-solution}
  q = (1-D)^{-1} \mathrm{diag}(P^{in}/F) q^{in}~,
\end{equation}
\end{linenomath}
where the inverse of $1-D$ can be shown to exist as Neumann series $(1-D)^{-1} =
\sum_{k=0}^{\infty} D^k$, since the absolute value of each eigenvalue of $D$ is
smaller than $1$, if there is at least one bus with a positive power injection
in each connected component, similarly to~\cite{achayuthakan2010}. While,
therefore, the method is formally applicable also in the presence of loop flows,
the interpretation of the resulting flow attribution still remains to be
investigated.


To apply flow tracing to the 4-bus example shown in Fig.~\ref{fig:4-node-num}
one calculates the in-partition $q^{in}$ (Eq.~(\ref{eq:qin-eu-num})), nodal flows
$F$, power injections $P^{in}$ and matrix $1-D$
\begin{linenomath}
\begin{IEEEeqnarray}{rCl}
  F &=& \begin{pmatrix} 11.4 & 1.9 & 2.9 & 2.8 \end{pmatrix}~,\\
  P^{in} &=& \begin{pmatrix} 11.4 & 0.9 & 0.5 & 0 \end{pmatrix}~,\\
  1-D &=&
  \begin{pmatrix}
    1        & 0        & 0 & 0 \\
    -1/1.9   & 1        & 0 & 0 \\
    -1.8/2.9 & -0.6/2.9 & 1 & 0 \\
    -1.9/2.8 & 0        & -0.9/2.8 & 1
  \end{pmatrix}~.
\end{IEEEeqnarray}
\end{linenomath}
By evaluation of Eq.~(\ref{eq:flow-tracing-solution}), one then finds
\begin{linenomath}
\begin{equation}
  q \approx
  \begin{pmatrix}
    0.921 &  0.0 &  0.000 &  0.0 &  0.079 \\
    0.485 &  0.0 &  0.000 &  0.0 &  0.515 \\
    0.672 &  0.0 &  0.172 &  0.0 &  0.156 \\
    0.841 &  0.0 &  0.055 &  0.0 &  0.104
  \end{pmatrix}~,
\end{equation}
\end{linenomath}
where each column corresponds to the share of each nodal flow associated with a
component, and consequently also to the share on the out-going lines. Since
buses 2 and 4 do not feed any power into the network, they do not contribute to
any flows. The shares of bus 1, which feeds all buses directly, are strongest at
bus 1 and 4, while at bus 2 a strong in-flow by imports dilutes the share of bus
1. Power entering the network as imports is present at every bus, making up
$15.6$\% of the nodal flow through bus 3 and $10.4$\% of the nodal flow at bus 4.
The generation of bus 3 only appears at buses 3 and 4.

Since the power loss happens on the links it is natural to attribute a loss
\(q_{n,\alpha} \chi_{n \to m}\) to entity \(\alpha\), f.ex. according to flow
tracing imported power leads to a loss of $q_{2,I} (F^{out}_{2\to 3} -
F^{in}_{2\to 3}) = 0.515 \cdot \SI{0.1}{GW}$ in line $2\to 3$. Substituting
\(F^{out}_{m \to n} = F^{in}_{m \to n} + \chi_{m \to n}\) in
Eq.~(\ref{eq:alpha_flow_conservation})
\begin{linenomath}
\begin{equation}
q_{n,\alpha}^{in} P_{n}^{+} + \sum_{m} q_{m,\alpha} F^{in}_{m \to n}
= q_{n,\alpha} \left( P_{n}^{-} + \sum_m \chi_{n\to m} \right) + q_{n,\alpha} \sum_{m} F^{in}_{n\to m}~,
\end{equation}
\end{linenomath}
reveals by comparing the structure again to Eq.~\eqref{eq:alpha_flow_conservation}
that this loss allocation scheme is equivalent to treating a loss on line \(n
\to m\) as an additional load at the out-flowing bus \(n\) combined with flow
tracing on the inflows indiscriminately, the procedure Bialek introduced as net
flows~\cite{bialek1996}.

In summary, for a given flow pattern \(F_{n\to m}\) and a fixed attribution of
the generated power \(q_{n,\alpha}^{in} P_{n}^{+}\) to a set of components
\(\alpha\) the flow tracing algorithm yields the attribution of all flows along
the links \(q_{n,\alpha}F_{n\to m}\) and the attribution of the power flowing
into the consuming nodes \(q_{n,\alpha}P_{n}^{-}\).

Note that we are able to invert the injection pattern and flow graph
consistently by switching the signs \(P_{n}^{\pm}\to P_{n}^{\mp}\) and the flows
\(F^{in/out}_{n \to m} \to F^{out/in}_{m \to n}\). This procedure allows a given
out-partition to be considered as the input for the flow tracing algorithm,
which then assigns shares of the power flow and injected power at the source
node according to this partition.

\subsection{A measure of transmission line usage}
\label{sec:usage}

The flow tracing method as displayed in the last section refers to the
application to a single flow pattern. However, for the analysis of complex
modern electricity systems, one rather has to consider whole time series of
fluctuating injection and flow patterns taking place on the underlying power
grid. The application of the flow tracing method then yields a time series
\(\left( F_{l}(t), \{q_{l,\alpha}(t)\} \right)\), containing the power flows
\(F_{l}(t)\) and the respective shares \(q_{l,\alpha}(t)\) assigned to the
components \(\alpha\) for each link \(l\). In order to derive the respective
grid usage over the whole time series, this information has to be integrated
into a suitable transmission capacity usage measure. Fig.~\ref{fig:usagestats}
below illustrates the need for such a non-trivial measure in a realistic
example: Consider, for instance, in cyan the shares associated with onshore wind
on a specific line; while the shares shown as small dots vary strongly over
time, their conditional averages,
\begin{linenomath}
\begin{equation}
  h_{l,\alpha}(F) = \left\langle\,q_{l,\alpha}(t)\,\right\rangle_{\{t | F_l(t) = F\}}~,
\end{equation}
\end{linenomath}
depend smoothly on the absolute line flow at which the average is taken. The
right plot in Fig.~\ref{fig:usagestats} shows a line where on average all the
power transmitted at a low line-loading is traced back to conventional
generators, while in hours with a high line-loading on- and offshore wind
contribute nearly all the power. Such a striking difference should be accounted
for as relevant information by an adequate usage measure. In the following we
briefly review such a capacity usage measure introduced in~\cite{tranberg2015}.

Their central idea is that the transmission line capacity of a small increment
between \(\mathcal{K}\) and \(\mathcal{K}+d\mathcal{K}\) is only used by flows
$F(t) > \mathcal{K}$ and, thus, the usage share of a component for this capacity
increment is determined only from those as
\begin{linenomath}
\begin{equation}
  w_{l,\alpha}(\mathcal{K}) = \left\langle\,q_{l,\alpha}(t)\,\right\rangle_{\{t | F_l(t) > \mathcal{K}\}}~.
\end{equation}
\end{linenomath}
Mind ``$>$'' in the subscript. The capacity of the whole line
$\mathcal{K}^T_{l,\alpha}$ can then be split for the individual components
$\alpha$ by summing all increments to
\begin{linenomath}
\begin{equation}
  \label{eq:usage-measure}
  \mathcal{K}^T_{l,\alpha} = \frac{\mathcal{K}^T_l}{\max_t F_l(t)} \int_{0}^{\max_t F_l(t)} \left\langle\,q_{l,\alpha}(t)\,\right\rangle_{\{t | F_l(t) > \mathcal{K}\}} \, \mathrm{d}\mathcal{K}~.
\end{equation}
\end{linenomath}
The proportional factor in front of the integral accounts for the fact that
$w_{l,\alpha}(\mathcal{K})$ vanishes at the maximum flow by sharing the
remaining security margin $\mathcal{K}^T_{l,\alpha} - \max_t F_l(t)$
proportionally, since it is important to all users of the capacity in an
analogous way to the actually used capacity. Nevertheless, depending on the
details of the system under investigation other schemes are possible.

\section{Flow tracing applied to a 118-bus electricity network model}
\label{sec:application}

\begin{figure}
\makebox[-3pt][l]{\includegraphics[width=\linewidth]{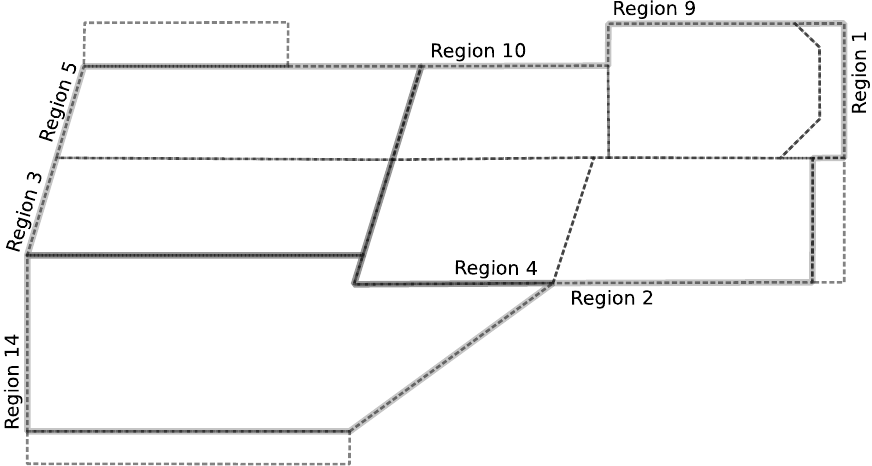}}
\includegraphics[width=\linewidth]{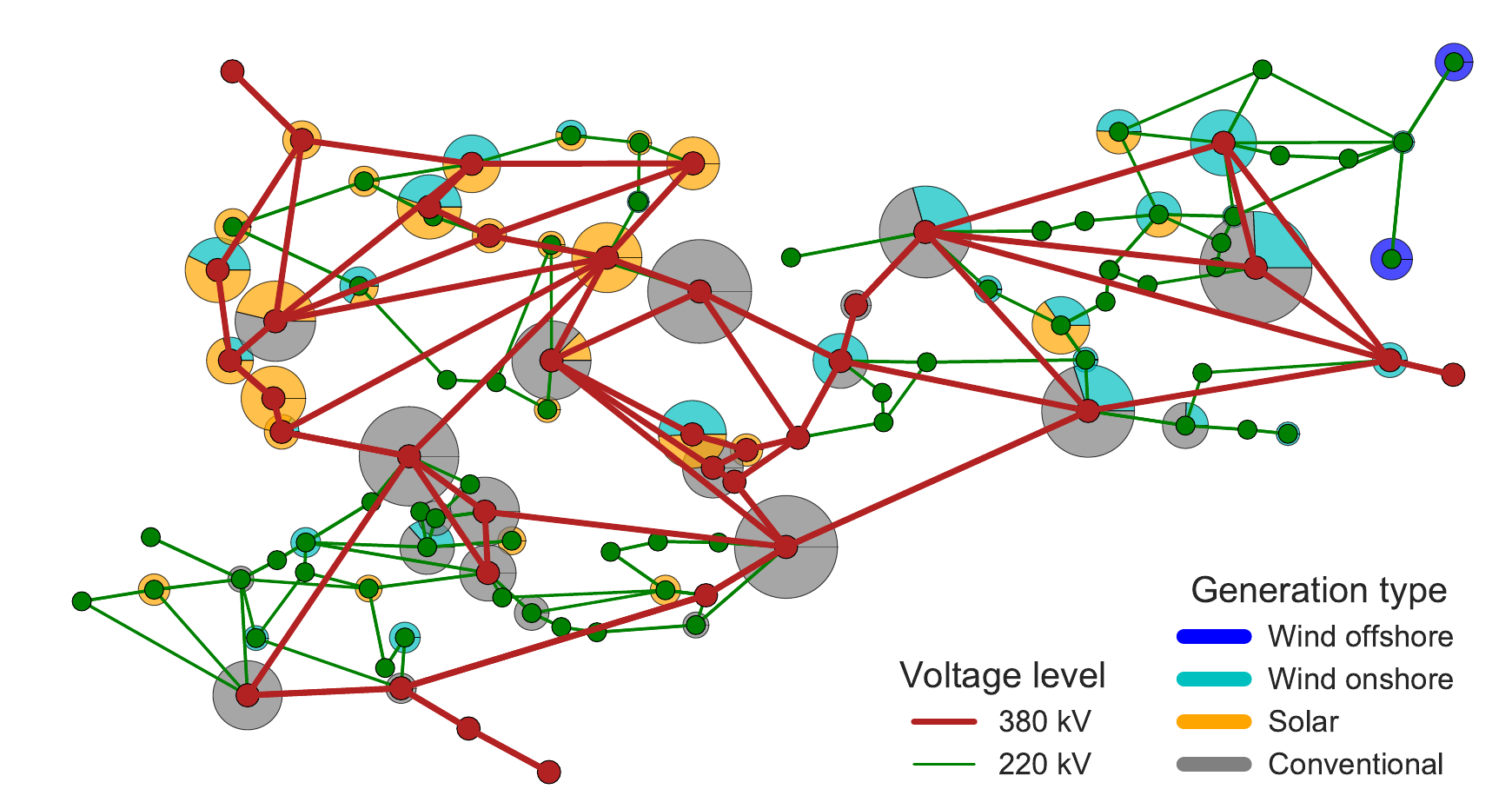}
\caption{Scenario 2023B of the 118-bus transmission
    expansion benchmark case with renewable generation capacities from
    \cite{barrios2015}. In the background the relative composition of
    the generation capacities of each region are indicated.}
\label{fig:barrios-model}
\end{figure}

For demonstrating the application of the reformulated flow tracing methodology, we
briefly introduce an electricity system model that has been developed
as a benchmark for transmission expansion methods. The IEEE 118 bus
network model has been geographically embedded and augmented by
attaching specific loads and conventional as well as renewable
generators to the system by Barrios et al. at RWTH
Aachen~\cite{barrios2015}. The load curves and the renewable generation
availability span all hours in a model year. The geographic regions
have weather characteristics in line with the artificial TRY-Regions
of the so-called TRY reference data set of the German weather service
(DWD), which feature a higher solar capacity factor in the North-West
and a higher wind capacity factor in the East, where, in addition, an
offshore-wind region is located. The network topology and generation
capacities are shown in Fig.~\ref{fig:barrios-model}, while the
average generation and consumption of each region are included in
Fig.~\ref{fig:generation-consumption-summary}.

We use our electricity system modeling framework PyPSA~\cite{pypsa} to
determine the linear optimal power flow (LOPF), i.e. the dispatch of
the generators is solved by a convex linear optimization minimizing
the total cost based on the marginal costs of the conventional
generators and the spatially and temporally fluctuating availability
of renewable generation subject to meeting the load curve and the
transmission constraints in all hours. Once the generator dispatch has
been determined, the non-linear power flow is found by a standard
Newton-Raphson iteration. Several key figures of the optimization are
summarized in Table~\ref{tab:characteristic}.

\begin{table}
  \centering
  \caption{Characteristic figures of the LOPF solution in units of {GW}}
  \label{tab:characteristic}
  \begin{tabular}{rccccccc}
  \toprule
  & \multicolumn{4}{c}{Generation} & Load & Loss \\
  & wind off. & wind on. & solar & conv.           &      &     \\
  \midrule
  capacities & \(1.8\) & \(21.2\) & \(22.3\) & \(27.9\) & - & - \\
  mean & \(0.8\) & \(4.5\) & \(2.6\)  & \(10.9\) & \(18.6\)  & \(0.2\) \\
  std  & \(0.7\) & \(4.3\) & \(4.1\)  & \(5.9\)  & \(3.2\)   & \(0.16\) \\
  min  & \(0\)   & \(0\)   & \(0\)    & \(0.05\) & \(9.4\)   & \(0.03\) \\
  max  & \(1.8\) & \(21.0\) & \(19.7\) & \(25.9\) & \(26.3\) & \(0.9\) \\
  \bottomrule
  \end{tabular}
\end{table}

\subsection{Analyzing mean flow patterns}
\label{sec:meanflow}

\begin{figure}
  \includegraphics[width=\linewidth]{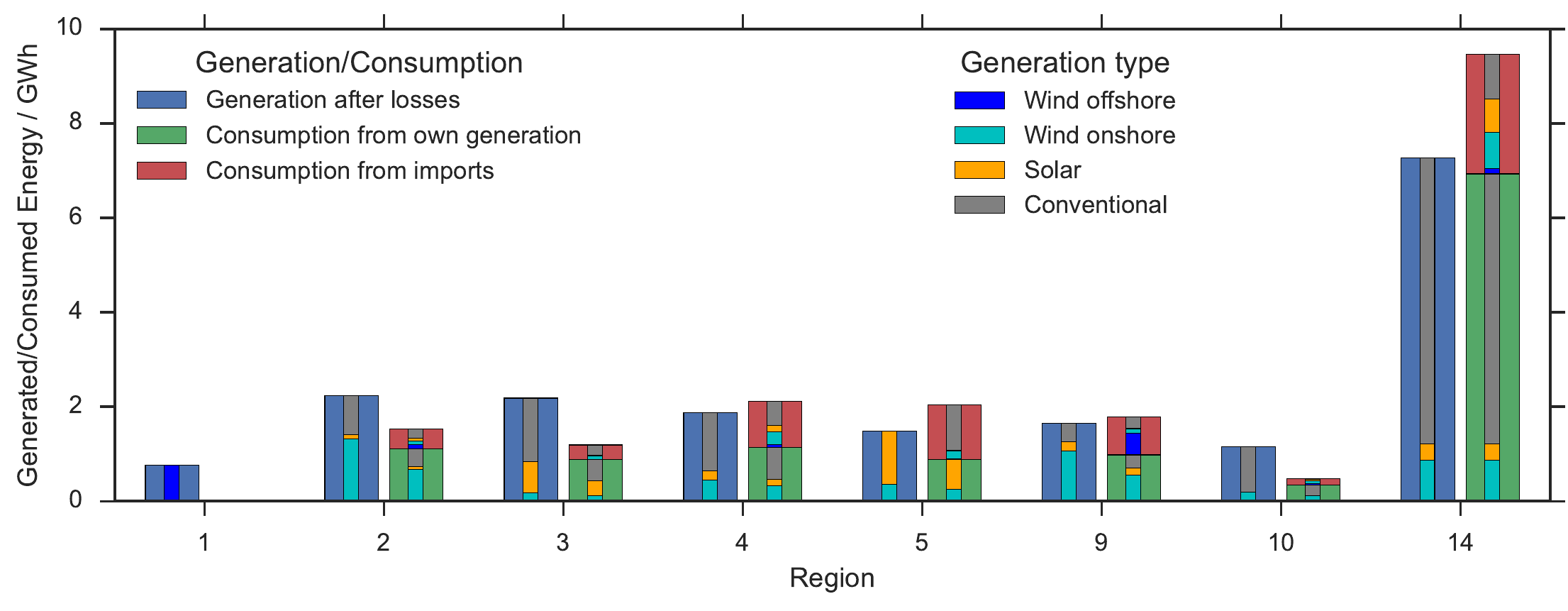}
  \caption{Comparison of the generated energy $\sum_{\beta} E_{\alpha,\beta}$
    (blue) in a region with its consumption from own generation
    $E_{\beta,\beta}$ (green) and from imports $\sum_{\alpha \neq \beta}
    E_{\alpha,\beta}$ (red) dissected with flow tracing on the in-partition
    Eq.~\eqref{eq:qin-regions}. The inset decomposition into energy per
    generation type has been calculated from the in-partition in
    Eq.~\eqref{eq:qin-regions-techs}.}
  \label{fig:generation-consumption-summary}
\end{figure}

\begin{figure}
  \floatbox[{\capbeside\thisfloatsetup{capbesideposition={left,bottom},capbesidewidth=0.4\textwidth}}]{figure}[\FBwidth]
  {\caption{Relative imports of a Region ($E_{\alpha,\beta} / \sum_{\alpha \neq
        \beta} E_{\alpha,\beta}$). For example, the value $0.6$ between Region
      $1$ and $9$ means that $60\%$ of the consumption from imports is covered
      by energy generated in Region $1$. 
    }\label{fig:generation-consumption-matrix}}
  {\includegraphics[width=0.6\textwidth]{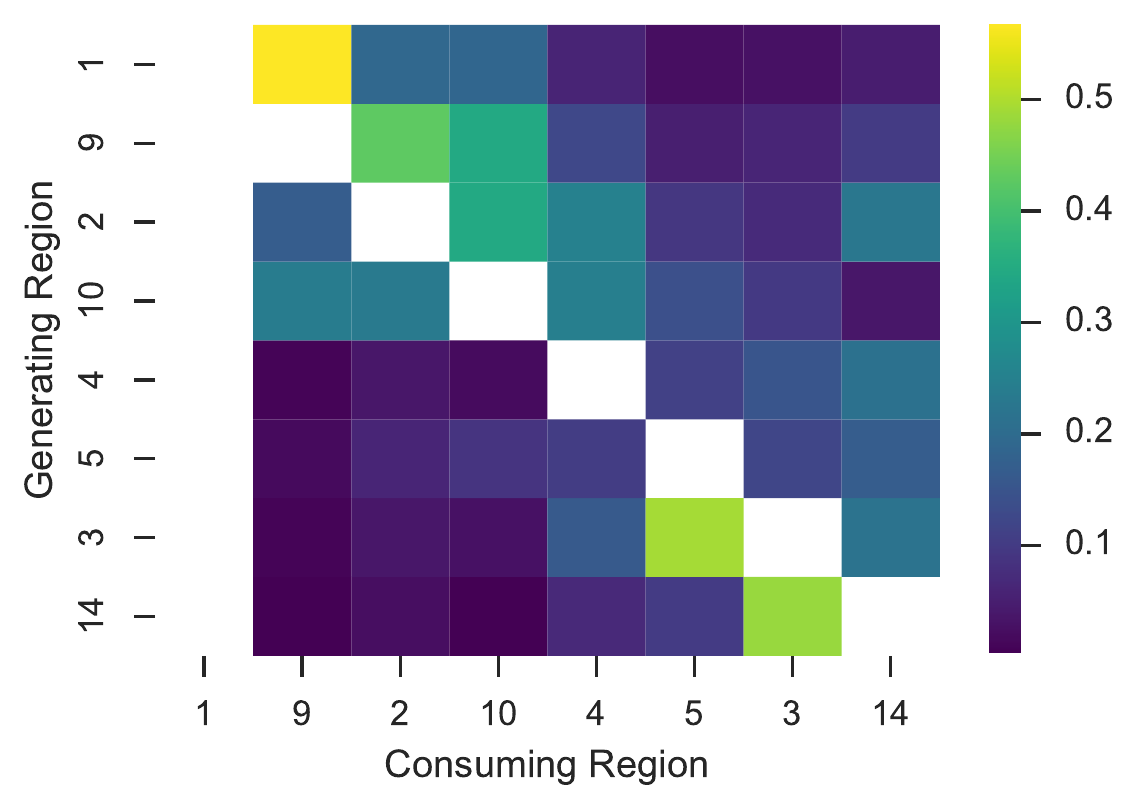}}
\end{figure}

To adopt the flow tracing method one initially distinguishes the injections per
region by choosing an in-partition
\begin{linenomath}
\begin{equation}
  q^{in}_{\alpha,n} = \delta_{\alpha,n} :=
  \begin{cases}
  1 & \text{ for node $n$ in region $\alpha$,} \\
  0 & \text{ else}
  \end{cases}~.
  \label{eq:qin-regions}
\end{equation}
\end{linenomath}

Using the flow tracing solution Eq.~\eqref{eq:flow-tracing-solution} we find the
partition \(q^{out}_{\alpha,n}(t)\) as the share of the energy consumed in bus
\(n\) that is generated in region $\alpha$, i.e. $q_{\alpha,n}(t)P_n^-(t)$. For
the total amount of energy from a region $\alpha$, we only need to correct for
the energy generated and consumed directly at bus $n$ given by $L_n-P^-_n$, if
bus $n$ also belongs to region $a$. The average inter-region flow from
region \(\alpha\) to region \(\beta\) then adds up to
\begin{linenomath}
\begin{equation}
  E_{\alpha,\beta} = \sum_{n \text{ in region } \beta} \left\langle \, q^{out}_{\alpha,n}
    \cdot P^-_{n} + \delta_{\alpha,n} \cdot (L_{n} - P^-_{n}) \, \right\rangle_{t}~.
  \label{eq:avg-flows}
\end{equation}
\end{linenomath}

These flows are illustrated in Fig.~\ref{fig:generation-consumption-summary} and
\ref{fig:generation-consumption-matrix}.

If you ignore the inset decomposition about renewables and focus on the outer
blocks in blue, red and green for now,
Fig.~\ref{fig:generation-consumption-summary} compares the net generated energy
$\sum_b E_{\alpha,\beta}$ to the consumed energy $\sum_\alpha E_{\alpha,\beta}$
in each region. The consumption has been decomposed into two parts which
are covered by local production and by imports. In contrast to summing up the
generation independent of flow tracing as $\sum_{n \text{ in region } \alpha}
G_n$, the small losses of about \(5\%\) for the energy generated in the offshore
Region~\(1\) and about \(2\%\) for the other regions have automatically been
netted away by considering directly the consumed energy.

The full benefit of using flow tracing for the average flow statistics becomes
only clear once we distinguish also between different generation types. We use
the components $\left\{(\alpha,\tau) | \alpha\in \text{regions},
  \tau\in\{\mathrm{wind_{on}}, \mathrm{wind_{off}}, \mathrm{solar},
  \mathrm{other}\}\right\}$ and extend the in-partition from
Eq.~\eqref{eq:qin-regions} to
\begin{linenomath}
\begin{equation}
  q^{in}_{(\alpha,\tau),n}(t) = \delta_{\alpha,n} G_n^\tau(t)/\sum_{\tau'} G_n^{\tau'}~,
  \label{eq:qin-regions-techs}
\end{equation}
\end{linenomath}
while Eq.~\eqref{eq:avg-flows} is adapted by substituting
$\alpha\to(\alpha,\tau)$. The resulting measure $E_{(\alpha,\tau),\beta}$ yields
the decomposition in generation types shown in
Fig.~\ref{fig:generation-consumption-summary}.

Regions with only one or two types of generation capacities in the
studied network model usually import a generation mix that is far more
balanced. This can be observed, for instance, in Region \(5\), which
only generates solar and wind energy, but consumes nevertheless more
than a third of conventionally generated energy and Region \(14\) with
mostly conventional generation capacities importing also a significant
amount of energy from renewable generation. It is also found that the
energy generated by offshore wind in Region \(1\) is mainly consumed
(to \(56\%\)) in the adjacent Region~\(9\) and only a tiny amount of
\(5\%\) reaches the remote Region~\(14\).

To study the spatial pattern on imports and exports in more detail, we decompose
the imports of each region further into the partial flows originating from each
of the other regions in Fig.~\ref{fig:generation-consumption-matrix}. The order
of the regions is chosen from the North-East to the South-West highlighting two
local clusters between regions \(1\), \(9\), \(2\) and \(10\) and between regions \(5\),
\(3\) and \(14\). Region \(4\) has a status of its own, since it receives most
of its imports from the north-eastern cluster, while it exports to the
south-western cluster. The high-load Region \(14\) satisfies also about a fourth
of its imports from regions \(2\) and \(9\) outside of its own cluster. This
indicates a net flow from the North-East to the South-West not unlike the German
situation of wind energy surpluses in the North-East flowing to the
load-intensive South and West.

Note that while, for simplicity, we studied the average energy flows,
all the partial flows are available as time-series retaining
correlations to important network characteristics. The following
section uses the correlations to line-loading for attributing
transmission line capacity.

\subsection{Attributing transmission capacity}
\label{sec:attributing}

In this section we will demonstrate the application of the line usage
measure reviewed in Sec.~\ref{sec:usage} to determine the transmission
capacity that is attributed to the four generation types on each
link. Extending the investigations in~\cite{tranberg2015}, we will
then compare the results to several other allocation measures. In
contrast to specific cost allocation models as
f.ex. Soares~et~al.~\cite{soares2015} propose for pricing distribution
grid capacities, our focus lies on improving the underlying usage
measure, in particular by incorporating correlations to the absolute
value of the power flows as detailed in Sec.~\ref{sec:usage}.

Usage shares of the transmission lines for generation types \(\tau\) are
captured by the line-flow partition \(\left\{ q_{l,\tau}(t) \right\}\) which
results from flow tracing on an in-partition
\begin{linenomath}
\begin{equation}
  q^{in}_{n,\tau}(t) = G^{\tau}_{n}(t) / \sum_{\tau''} \left( G^{\tau'}_n(t) \right)~,
  \label{eq:qin-techs}
\end{equation}
\end{linenomath}
based on the hourly energy generation mix \(G^{\tau}_{n}(t)\).

\begin{figure}
  \centering
  \begin{minipage}[t]{.42\linewidth}
    \centering
    \includegraphics[height=1.8in]{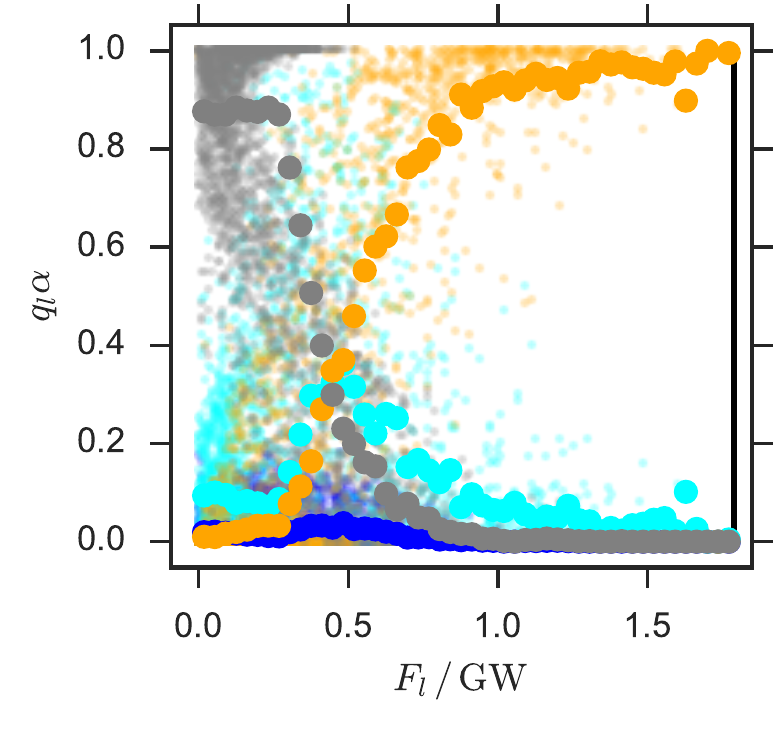}
  \end{minipage}
  \hfill
  \begin{minipage}[t]{.56\linewidth}
    \centering
    \includegraphics[height=1.8in]{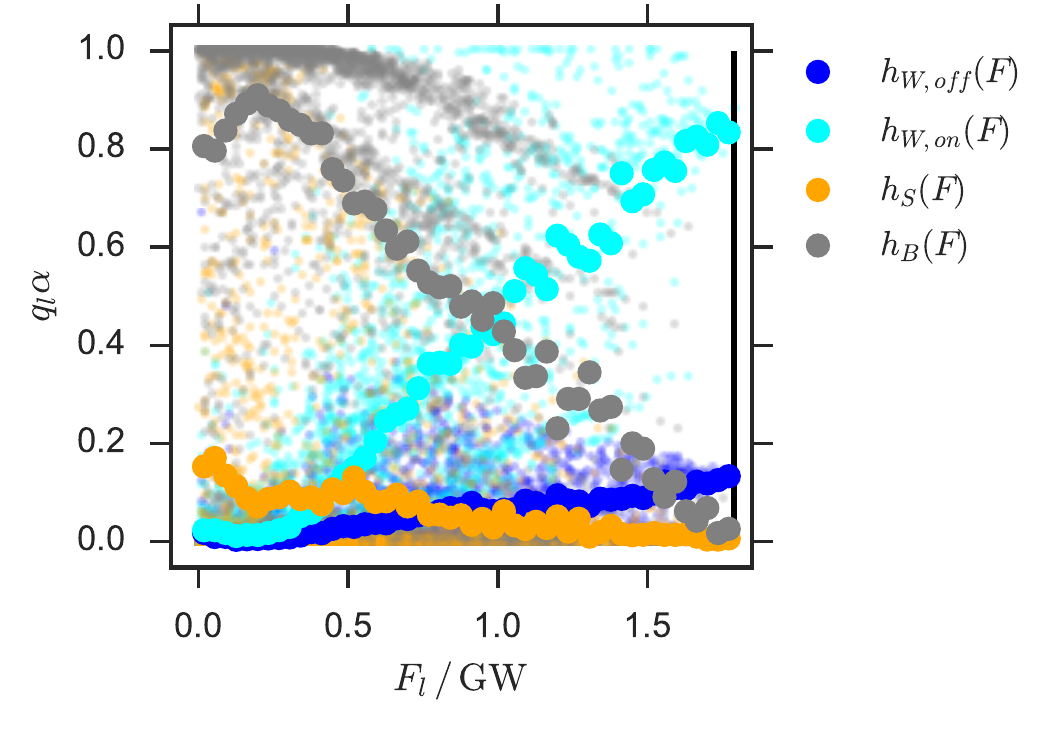}
  \end{minipage}%
  \caption{ Usage share of two single lines: The left power line
    connecting Region $3$ and Region $14$ on the left and the power
    line between Region $2$ and Region $14$ on the right, both
    highlighted in
    Fig.~\ref{fig:transmission-caps-by-generation-and-link}. Data
    points $(F_l(t), q_{l,\alpha}(t))$ for all hours and the
    conditional average shares $h_{l,\alpha}(F_l)$.  }
  \label{fig:usagestats}
\end{figure}

These shares vary significantly with the flow on a power line. For
both lines shown in Fig.~\ref{fig:usagestats} conventionally generated
power has a high share only in hours with low amounts of flow. In
hours with a high line-loading the line in the West is mainly occupied
by energy traced back to solar panels, while the line in the East
carries mostly wind power.

\begin{figure}
  \centering
  \makebox[-.5em][l]{\includegraphics[width=\linewidth]{map}}
  \includegraphics[width=\linewidth]{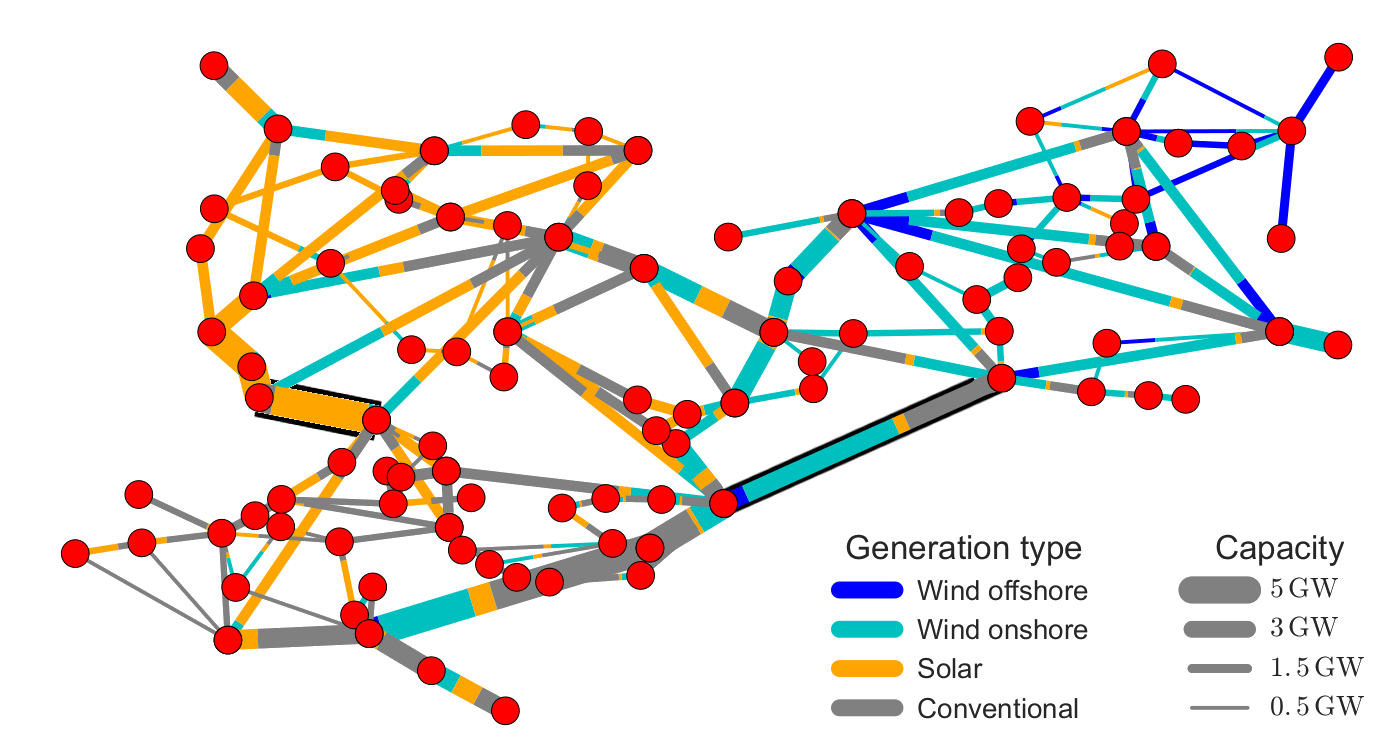}
  \caption{ Line capacities $\bar{\mathcal{K}}^T_{l,\alpha}$
    attributed to the four generation types $\alpha$ for each link $l$
    throughout the network of the benchmark case. A clear east-west
    separation is visible: In the East the high wind generation
    capacity is reflected in lines primarily loaded by wind energy,
    while in the West solar energy is dominating. The links
    highlighted by a black frame have been discussed in detail in
    Figure \ref{fig:usagestats}.}
  \label{fig:transmission-caps-by-generation-and-link}
\end{figure}

If one neglected this correlation for assessing the usage of the
latter eastern power line, one would find that the total amount of
conventional power is about a third higher than the amount of wind
power flowing through it and, thus, would conclude that the costs of
the power line should be split in the same proportion. Instead, the
reviewed usage measure from Sec.~\ref{sec:usage} gives a higher weight
to the shares with high line loads. To compare attributed transmission
capacities in line with transmission cost drivers length and capacity,
we multiply it by length \(\bar L_l\) as
\begin{linenomath}
\begin{equation}
  \label{eq:usage-measure_2}
  \mathcal{\bar K}^T_{l,\alpha} = \frac{\bar L_l \mathcal{K}^T_l}{\max_t F_l(t)} \int_{0}^{\max_t F_l(t)} \left\langle\,q_{l,\alpha}(t)\,\right\rangle_{\{t | F_l(t) > \mathcal{K}\}} \, \mathrm{d}\mathcal{K}~.
\end{equation}
\end{linenomath}
Similarly we understand the total transmission capacity of the network
to be given by \(\bar{\mathcal{K}}^T = \sum_l \mathcal{K}^T_l \bar L_l\) in
units of MW km.  The evaluation of the measure for the eastern line in
Fig.~\ref{fig:usagestats} attributes \(54\%\) to onshore wind and only
\(34\%\) to conventionally generated power.

The attributed capacities of all the transmission lines in the network
are shown schematically in
Fig.~\ref{fig:transmission-caps-by-generation-and-link}. Power from
onshore wind turbines takes up most of the capacity in the East, while
power generated by solar panels is attributed the transmission
capacity in the West. This separation mirrors the distribution of the
generation capacities (cf.~Fig.~\ref{fig:barrios-model}). Renewable
generation is attributed a share of the transmission capacity that is
disproportionately high compared to the average energy generation mix,
given in Table~\ref{tab:characteristic}. In Region \(14\), for instance,
where only few renewable generation capacities are located,
significant amounts of transmission capacity are attributed to solar
and wind generation.

We finally compare the flow tracing based usage measure in
Eq.~(\ref{eq:usage-measure_2}) with several alternative allocation
mechanisms for transmission capacity:

\emph{Average power injection} splits the transmission capacity of the
network \(\bar{\mathcal{K}}^T\) in proportion to the amount of injected
power of each generator \((n, \tau)\), i.e.
\begin{linenomath}
\begin{equation}
\mathcal{M}_{\alpha,\tau}^{(1)} =
\left(
  \frac{\sum_{\text{$n$ in region $\alpha$}} \expect{P^+_{n,\tau}}_t}
       {\sum_m \expect{P^+_{m,\tau}}_t}
\right)
\bar{\mathcal{K}}^T
~,
\end{equation}
\end{linenomath}
where \(P^+_{n,\tau}(t)\) is the power injected by a generation type
\(\tau\) at bus \(n\). This scheme corresponds to the widely used postage
stamp pricing mechanism.

\emph{Average power injection with topological correction} adjusts
\(\mathcal{M}_{\alpha,\tau}^{(1)}\) with an additional factor penalizing
remote locations, where the generators on average have to send their
energy farther through the network than from a central bus.
\begin{linenomath}
\begin{equation}
\mathcal{M}_{\alpha,\tau}^{(2)} =
\left(
     \frac{\sum_{\text{$n$ in region $\alpha$}} \expect{P^+_{n,\tau}}_t \bar{D}_n}
          {\sum_m \expect{P^+_{m,\tau}}_t \bar{D}_m}
\right)
\bar{\mathcal{K}}^T
~.
\end{equation}
\end{linenomath}
\(\bar{D}_n\) is the average graph distance of the bus \(n\), which is the
mean distance to the other buses~\cite{newman2010-book}.

\emph{Flow tracing mean usage} weights the attributions with a
distribution determined from the average line loading of each
generation type and region.
\begin{linenomath}
\begin{equation}
\mathcal{M}_{\alpha,\tau}^{(3)} = \left( \frac{\sum_l \expect{q_{l,\alpha,\tau} F_l}_t \bar L_l}{\sum_{l'} \expect{F_{l'}}_t \bar L_{l'}} \right)
\bar{\mathcal{K}}^{T}
\end{equation}
\end{linenomath}
It is similar to previously proposed pricing schemes based on flow
tracing~\cite{bialek1997,bialek1998}.

Finally, \emph{Flow tracing usage measure} distributes the capacity of
each line by the usage measure from Eq.~(\ref{eq:usage-measure_2}).
\begin{linenomath}
\begin{equation}
\mathcal{M}_{\alpha,\tau}^{(4)} = \sum_l \bar{\mathcal{K}}^T_{l,\alpha,\tau}
\end{equation}
\end{linenomath}

\begin{sidewaysfigure}
  \begin{minipage}[t]{.79\linewidth}
    \centering
    \includegraphics[width=\linewidth]{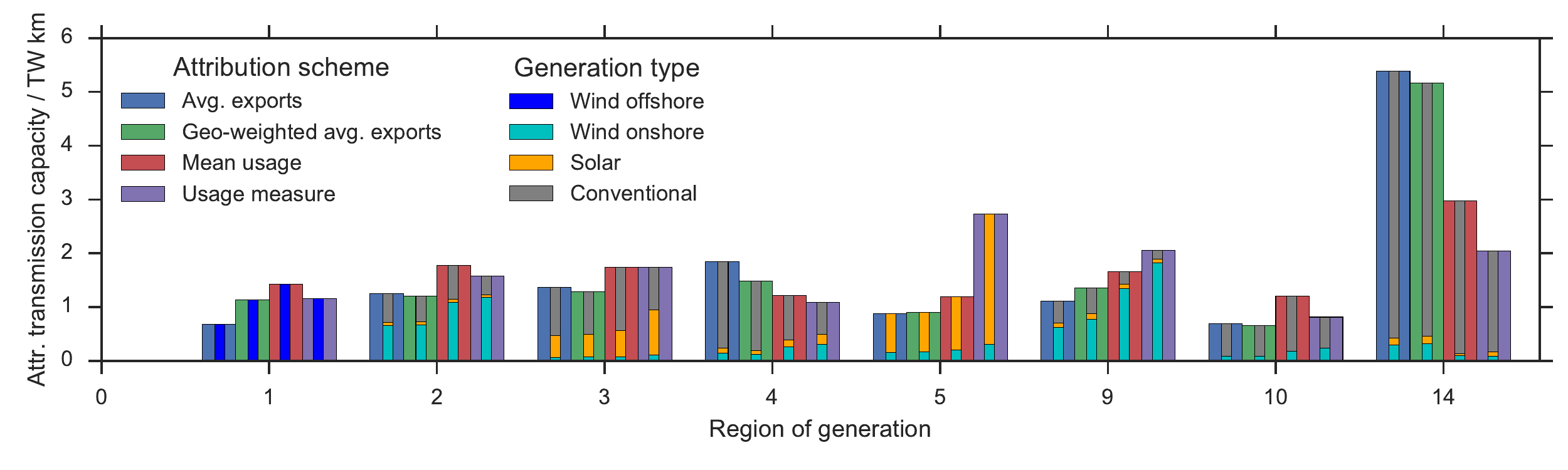}
  \end{minipage}
  \hfill
  \begin{minipage}[t]{.147\linewidth}
    \centering
  \raisebox{.14\linewidth}{\includegraphics[width=\linewidth]{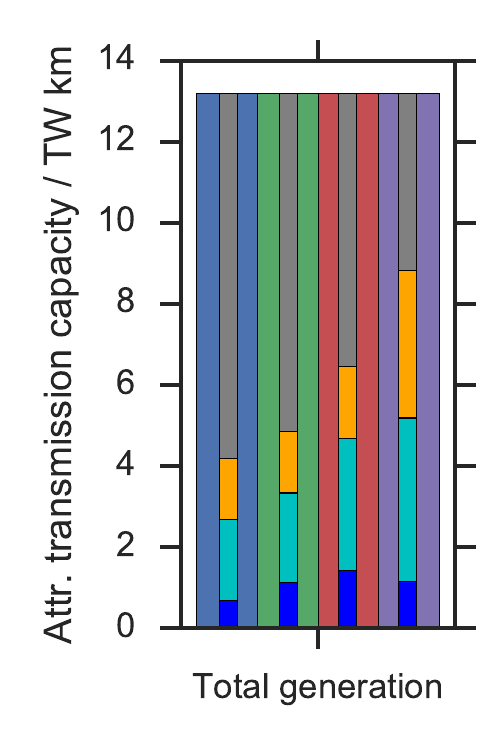}}
  \end{minipage}%
  \caption{
    Line capacities of the overall electricity system attributed to
    the generators of each Region using four different assignment
    schemes explained in more detail in the text, including the flow
    tracing usage measure from Eq. (\ref{eq:usage-measure_2}).
  }
  \label{fig:capacity-transmission-scheme-comparison}
\end{sidewaysfigure}

The four measures \(\mathcal{M}_{\alpha,\tau}^{(1)}\) to
\(\mathcal{M}_{\alpha,\tau}^{(4)}\) for the 118-bus network model are
illustrated in
Figure~\ref{fig:capacity-transmission-scheme-comparison}.

The geometry factor which distinguishes
\(\mathcal{M}_{\alpha,\tau}^{(2)}\) from
\(\mathcal{M}_{\alpha,\tau}^{(1)}\) only has a marginal effect on the
allocation, nevertheless it is still worth noting that the
modification is mostly in direction of the results of more elaborate
measures.

For most regions and generation types the simple measures
\(\mathcal{M}_{\alpha,\tau}^{(1/2)}\) agree quite well with the flow
tracing based measures \(\mathcal{M}_{\alpha,\tau}^{(3/4)}\).
Incorporating the actual shares of the line loading by flow tracing
turns out to have the largest effect for Region \(14\) which has a large
consumption and exclusively conventional generation. Most of the power
that is generated in Region \(14\) is consumed within few line
kilometers so that actual network transmission is kept to a minimum,
although the total power injected into the network is very high. This
indicates that an average distance to the load centers instead of to
all the buses in equal weights might be a better measure.

The capacity attributed to Region \(14\) is further reduced by taking
the line-loading correlations into account, since its conventional
generation is mainly dispatched in times with low renewable generation
and, thus, also small overall flows. But the most striking adjustment
from \(\mathcal{M}_{\alpha,\tau}^{(3)}\) to
\(\mathcal{M}_{\alpha,\tau}^{(4)}\) is that the capacity attributed to
the solar generation in Region \(5\) doubles, which is due to the strong
correlation between line-loading and solar flows already visible in
the single line usage share of Fig.~\ref{fig:usagestats}. The same
effect can also be seen in the usage measure component for solar power
generated in Region \(3\), only that the overall capacity for Region \(4\)
balances out thanks to its conventionally dominated generation mix.

Overall, we find that wind generators are strongly affected by switching from a
postage stamp pricing mechanism to a flow tracing based one, since the volume of
wind energy in the network is often disproportionately high as a study by Brown
et al., based on marginal participation, has already pointed
out~\cite{brown2015}. Additionally taking the correlation between line-loading
and usage shares into account strongly impacts the capacities attributed to
solar generation. The choice of a suitable capacity allocation measure thus
depends on the range of system properties which should be represented. Whereas a
simple postage stamp method might cover average imports and exports of the
system participants, only more elaborate techniques based on flow allocation are
able to incorporate the correlations and patterns emerging from the fluctuating
imports and exports in large-scale electricity systems with a high share of
renewable generation.

\section{Conclusion and Outlook}
\label{sec:concl}

Flow tracing is a well-known method to dissect the power flows on a
network according to shares attributed to the network-injecting source
nodes~\cite{bialek1996,kirschen1997}. Such an attribution of power
flows and thus network usage has been proposed as an essential component for a fair
allocation of both operational (for instance losses or stability
measures) and grid infrastructure costs~\cite{bialek2003, bialek2004-2, consentec06, tranberg2015}.
In the present contribution we show how a reformulation of the flow tracing
method serves as powerful tool set to analyze the complex spatio-temporal
patterns of generation, consumption and power flows in interconnected
large-scale electricity systems, in particular those with a high share of
renewable generation. At the point of injection into the network, the power flow
can straightforwardly be assigned to a specific geographical location, mix of
different generation technologies or any other attribution of interest.
Following the composition of ingoing flow from the net generators through the
network, the algorithm yields in an intuitive way the respective shares of the total power flow and
of the outflow to the net consumers.

The potential of this method are illustrated in the context of the Scenario
2023B of the 118-bus transmission benchmark case with renewable
capacities~\cite{barrios2015}. We dissect the power flows into components
associated with the geographical origin and generation technology (wind
offshore/onshore, solar, conventional), yielding a selection of measures about
the respective transmission system usage and the corresponding import-export
relations between the system nodes. The need of incorporating relevant
correlations in the aggregation procedure from full high-dimensional results to
a lower-dimensional expression is discussed by comparing different transmission
capacity usage measures assigned to the geographical regions of the benchmark
network.

The discussion in the present paper suggests future work on different levels.
From a technical point of view, it will be interesting to transfer the idea of
more general in-partitions to alternative methods of flow
allocation~\cite{chen2016,lima2004,lima2009}. First steps in this direction have
been taken by~\cite{brown2015} for allocation methods based on power transfer
distribution factors, but a rigorous discussion is still lacking in the
literature. In the present paper we focus on illustrating the versatility of the
reformulated flow tracing method by considering a well-defined and fully open
benchmark system. We are confident that an approach as discussed here will be
integrated into a state of the art tool set of system analysis and applied to
scenarios of the current and future electricity system, helping to reveal
patterns in the intricate interactions in this highly complex multi-level
system. As shown for the test case, it is then challenging to select and adapt
suitable measures from the the multitude of information which can be yielded
using this technique.

\section{Acknowledgments}
We thank Tom Brown for constructive comments and discussions. The
project underlying this report was supported by the German Federal
Ministry of Education and Research under grant no.~03SF0472C. The
responsibility for the contents lies with the
authors. M.S. acknowledges support from Stiftung Polytechnische
Gesellschaft Frankfurt and the Carlsberg Foundation.

\section*{References}
\bibliography{flowtracing}

\end{document}